\documentclass[amsmath,amssymb,showpacs,twocolumn,superscriptaddress,pr]{revtex4-1}
\usepackage{graphicx,bm,color,subfigure}

\begin{document}

\title{$d+id$ chiral superconductivity in a triangular lattice from trigonal bipyramidal complexes}

\author{Chen Lu}
\thanks{These two authors contributed equally to this work.}
\affiliation{School of Physics, Beijing Institute of Technology, Beijing 100081, China}

\author{Li-Da Zhang}
\thanks{These two authors contributed equally to this work.}
\affiliation{School of Physics, Beijing Institute of Technology, Beijing 100081, China}

\author{Xianxin Wu}
\affiliation{Institut f\"ur Theoretische Physik und Astrophysik,
	Julius-Maximilians-Universit\"at W\"urzburg, 97074 W\"urzburg, Germany}
\affiliation{Institute of Physics, Chinese Academy of Sciences, Beijing 100190, China}
\author{Fan Yang}
\email{yangfan\_blg@bit.edu.cn}
\affiliation{School of Physics, Beijing Institute of Technology, Beijing 100081, China}

\author{Jiangping Hu}
\email{jphu@iphy.ac.cn}
\affiliation{Institute of Physics, Chinese Academy of Sciences, Beijing 100190, China}
\affiliation{Collaborative Innovation Center of Quantum Matter, Beijing 100871, China}
\affiliation{Kavli Institute of Theoretical Sciences, University of Chinese Academy of Sciences,
Beijing, 100190, China}

\begin{abstract}
We model the newly predicted high-$T_c$ superconducting candidates constructed by corner-shared trigonal bipyramidal complexes  with an effective three-orbital tight-banding Hamiltonian and investigate the pairing symmetry of their superconducting states driven by electron-electron interactions. Our combined weak and strong coupling based calculations consistently identify the chiral $d+id$ superconductivity as the leading pairing symmetry in a wide doping range with realistic interaction parameters. This pairing state has nontrivial topological Chern-number and can host gapless chiral edge modes, and the vortex cores under magnetic field can carry Majorana zero modes.
\end{abstract}



\maketitle


\section{Introduction}
It has been long dream for the physics community to understand the pairing mechanism of high-$T_c$ superconductivity (SC). In the past three decades, the synthesized superconductors families are the cuprates and the iron pnictides.  Based on the common electronic properties of these two superconductors family \cite{Johnston}, one of us proposed two basic principles to unify the understanding for both superconducting families \cite{Hu}, which can serve as guiding rules to search for high-$T_c$ superconductors. Firstly, the correspondence principle \cite{HuDing,DavisLee} requires that the short-range magnetic exchange interactions and the Fermi surfaces (FSs) act collaboratively to achieve high-$T_c$ SC and determine pairing symmetries. This principle provides an unified explanation for the origin of the $d$-wave and the $s$-wave pairing symmetries in the cuprates and iron-based superconductors respectively \cite{Hu}. Secondly, the selective magnetic pairing rule points out  that the SC is   induced by the magnetic exchange couplings caused by the superexchange mechanism through cation-anion-cation chemical bonds, but not by the  magnetic exchange couplings from the direct exchange  mechanism through cations $d$-$d$ chemical bonds . This rule explains why the cuprates formed by cation-anion octahedral complex with Cu$^{+2}$ $d^9$ filling configuration and the iron-pnictides formed by tetrahedral complex with Fe$^{+2}$ $d^6$ filling configuration are high-$T_c$ superconductors, as the quasi-two-dimensional electronic environments in these two structures are dominated by the $d$ orbitals with the strongest in-plane $d$-$p$ hybridization near Fermi energy. Based on the two principles, a third family of  high-$T_c$ superconductors  is predicted in Ref. \cite{YNO}, which are formed by cation-anion trigonal bipyramidal complexes with a $d^7$ filling configuration on the cation ions.

\begin{figure}[htbp]
\centering
\includegraphics[width=0.48\textwidth]{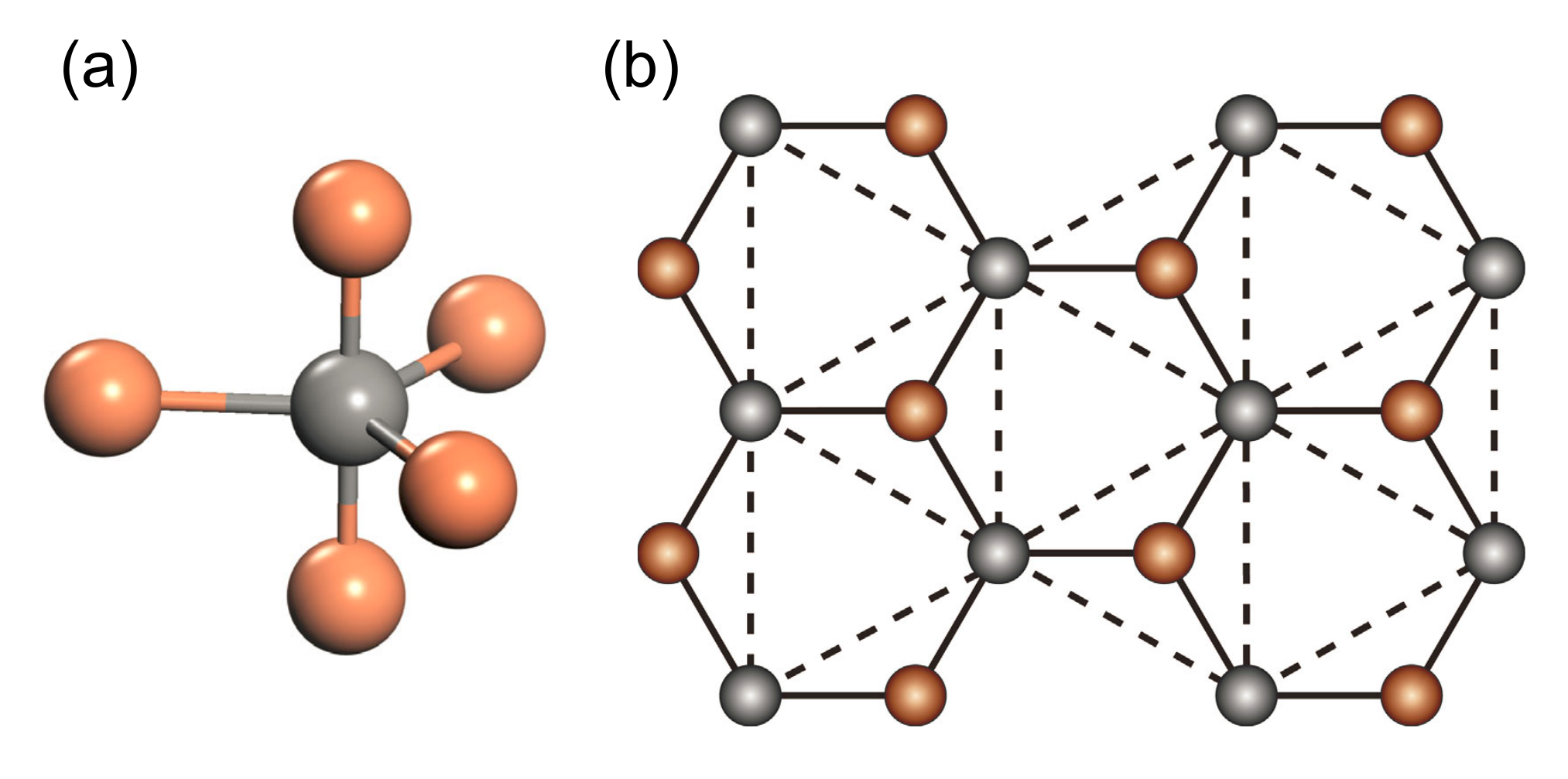}
\caption{The effective lattice structure of the trigonal bipyramidal complex YNiO$_3$. (a) The trigonal bipyramidal structure formed by the (brown) O anions around the (gray) Ni cation. (b) The two-dimension honeycomb lattice formed by the in-plane Ni cations and O anions. The Ni cations themselves form a triangle lattice.}\label{lattice}
\end{figure}

In the  material database, YMnO$_3$\cite{Yakel,Smolenskii}  can be considered as the simplest prototype of this structure. However,  Mn$^{3+}$ only hosts a $d^4$ filling configuration. To achieve the $d^7$ filling configuration,  we can theoretically consider YNiO$_3$  with the same structure as YMnO$_3$.  Here, Ni$^{3+}$ has a d$^7$ filling configuration. The effective lattice structure of YNiO$_3$ is shown in Fig. \ref{lattice}, where the O anions around the Ni cation form the trigonal bipyramidal structure, the in-plane Ni cations and O anions form the two-dimension honeycomb lattice, and the Ni cations themselves form a triangle lattice. Since its interlayer coupling is weak comparing with the intralayer one, we can ignore the third dimension, and regard YNiO$_3$ as a monolayer two-dimension material in the theoretical study. It's analyzed in Ref. \cite{YNO} that for the d$^7$ filling configuration of Ni atom in the trigonal bipyramidal complex, the low energy degree of freedom near the Fermi level is dominated by the d$_{xy/x^2-y^2}$ orbitals, which hybridizes strongly with the $p$ orbitals of the in-plane oxygen atoms. Such structure is consistent with the above introduced second principle, and thus favors the formation of high-$T_c$ SC.

In this paper, we calculate the superconducting pairing symmetry of YNiO$_3$ in both weak and strong coupling limit. We first model the single layer of YNiO$_3$ with a effective three-orbital tight-band (TB) Hamiltonian. Then we investigate its pairing symmetry based on the Hubbard-Hund model of the system in the weak coupling limit. Our calculation based on the random phase approximation (RPA) identifies that, for the realistic interaction parameters, the singlet $d+id$ pairing state driven by the antiferromagnetic (AFM) spin fluctuations dominates over other pairing states and serves as the leading pairing symmetry within a rather wide doping range. Furthermore, we investigate the pairing symmetry of the system based on the effective $t-J$ model in the strong coupling limit, which yields consistent result with that of RPA in the weak coupling limit. To investigate the topological properties of such a $d+id$ pairing state, we calculate the Chern number and the edge spectrum of the system. As a result, we obtain a Chern number of 8, and consequently there are 8 topological protected chiral gapless Dirac modes on each edge. When Rashba spin-orbit coupling (SOC) is included, the vortex cores under magnetic field can carry Majorana zero modes, which can be equipped with topological quantum computation.

The rest of this paper is organized as follows. In Sec. II, we describe the effective three-orbital TB model for YNiO$_3$. In Sec. III, we study the superconducting pairing symmetry of the system in the weak coupling limit, and provide the phase-diagram of the system in different parameter spaces. In Sec. IV, we reinvestigate the problem in the strong coupling limit. In Sec. V, we calculate the topological invariant and the edge spectrum. Finally, in Sec. VI, a conclusion will be reached with some discussions.

\section{The TB Model}
In the lattice structure of YNiO$_3$ as shown in Fig. \ref{lattice}, the $d_{xz}$ and $d_{yz}$ orbitals of the Ni cations have the lowest energy and are only weakly coupled to O anions \cite{YNO}. Thus we can assume them to be fully occupied, and model the system with the following effective three-orbital TB Hamiltonian:
\begin{align}\label{hk}
h(\bm{k})=\left(\begin{array}{ccc}
h_{11}& h_{12} & h_{13} \\
      & h_{22} & h_{23} \\
      &        & h_{33} \\
\end{array}\right).
\end{align}
Here the orbital indices $1$, $2$, and $3$ denote $d_{z^2}$, $d_{xy}$, and $d_{x^2-y^2}$ orbitals, respectively, and the unshown matrix elements can be obtained from the shown ones by the Hermicity of $h(\bm{k})$.

Introducing $x=\frac{\sqrt{3}}{2}{k_xa_0}$, $y=\frac{1}{2}k_ya_0$, where the $a_0$ is the lattice constant, we can write the matrix elements of $h(\bm{k})$ as follows:
\begin{align}\label{hop}
h_{11}=&\epsilon_1+2s^{11}(\cos2y+2\cos x\cos y),            \nonumber\\
h_{12}=&2\sqrt{3}s^{12}_{2}\sin x\sin y
+2is^{12}_{1}(\sin 2y+\cos x\sin y)                       \nonumber\\
h_{21}=&h_{12}^{*}                                        \nonumber\\
h_{13}=&2s^{12}_{2}(\cos 2y-\cos x\cos y)
-2i\sqrt{3}s^{12}_{1}\sin x\cos y                        \nonumber\\
h_{31}=&h_{13}^{*}                                         \nonumber\\
h_{22}=&\epsilon_2+2s^{22}_{11}\cos 2y
+(s^{22}_{11}+3s^{22}_{22})\cos x \cos y                               \nonumber\\
h_{23}=&\sqrt{3}(s^{22}_{11}-s^{22}_{22})\sin x\sin y          \nonumber\\
&+2is^{22}_{12}(\sin 2y-2\cos x \sin y)                                       \nonumber\\
h_{32}=&h_{23}^{*}                                     \nonumber\\
h_{33}=&\epsilon_2+2s^{22}_{22}\cos 2y
+(3s^{22}_{11}+s^{22}_{22})\cos x \cos y
\end{align}
The hopping parameters in the above matrix elements obtained by least-square-root fitting of the above TB model to the DFT band structure are listed in Table \ref{hopping}.

\begin{table}[htbp]
\caption{The hopping parameters (in unit of eV) in the three-orbital TB model. The on-site energies are $\epsilon_1=4.186455$eV, $\epsilon_2=2.765428$eV.}\label{hopping}
\begin{ruledtabular}
\begin{tabular}{cccccc}
$s^{11}$ & $s^{12}_{1}$ & $s^{12}_{2}$ & $s^{22}_{11}$ & $s^{22}_{22}$ & $s^{22}_{12}$ \\
\colrule
-0.1639 & 0.2063 & 0.0678 & 0.3147 & 0.1091 & 0.0388 \\
\end{tabular}
\end{ruledtabular}
\end{table}

\begin{figure}[htbp]
\centering
\includegraphics[width=0.48\textwidth]{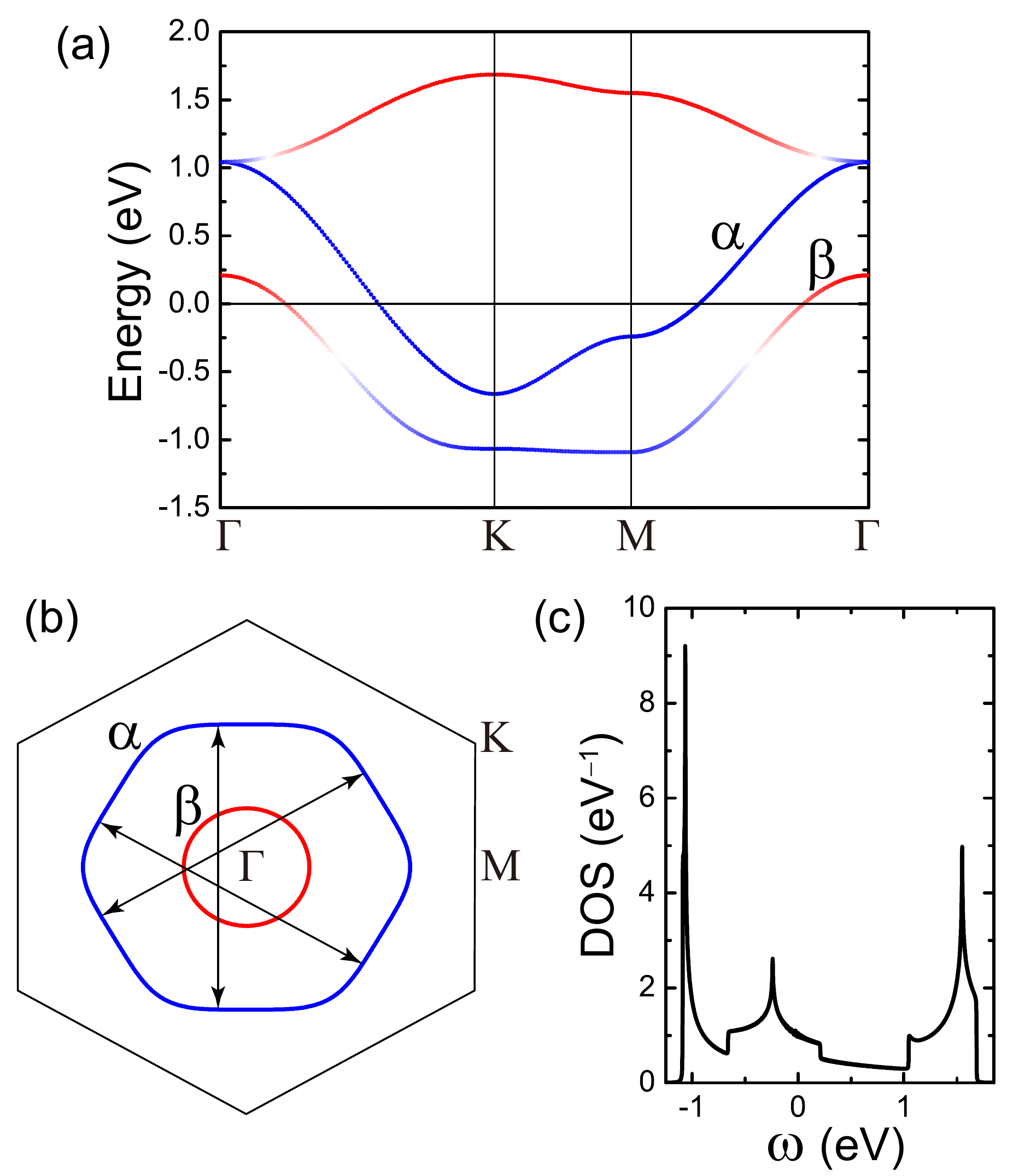}
\caption{(a) The effective band structure of the undoped YNiO$_3$ along the high-symmetric lines. The red parts of the bands denote those with the Ni-$d_{z^2}$ orbital character; the blue parts of the bands denote those with Ni-$d_{xy}$ and -$d_{x^2-y^2}$ orbital characters. The $\alpha$ and $\beta$ mark the two bands crossing the Fermi level. (b) The FS sheets of the undoped system corresponding to the bands $\alpha$ and $\beta$. (c) The DOS of the undoped system.}\label{bandfsdos}
\end{figure}

The band structure of the above TB model is shown in Fig. \ref{bandfsdos}(a). The chemical potential is $\mu_c=2.99$eV, which leads to a band filling of 3 electrons per unit cell. The two bands crossing the Fermi level are marked as $\alpha$ and $\beta$ in Fig. \ref{bandfsdos}(a), and the FS sheets are marked correspondingly in Fig. \ref{bandfsdos}(b). The orbital character of band $\alpha$ is Ni-$d_{xy}$ and -$d_{x^2-y^2}$, and the orbital character of band $\beta$ is $d_{z^2}$ as shown in Fig. \ref{bandfsdos}(a). The density of states (DOS) of the system is shown in Fig. \ref{bandfsdos}(c).

\section{Weak coupling limit: RPA approach}
In the weak coupling limit, we adopt the following Hamiltonian in our RPA calculations:
\begin{align}\label{model}
H&=H_{tb}+H_{int}\nonumber\\
H_{int}&=U\sum_{i\mu}n_{i\mu\uparrow}n_{i\mu\downarrow}+
V\sum_{i,\mu<\nu}n_{i\mu}n_{i\nu}+J_{H}\sum_{i,\mu<\nu}                   \nonumber\\
&\Big[\sum_{\sigma\sigma^{\prime}}c^{+}_{i\mu\sigma}c^{+}_{i\nu\sigma^{\prime}}
c_{i\mu\sigma^{\prime}}c_{i\nu\sigma}+(c^{+}_{i\mu\uparrow}c^{+}_{i\mu\downarrow}
c_{i\nu\downarrow}c_{i\nu\uparrow}+h.c.)\Big].
\end{align}
Here, the $U$, $V$, and $J_H$ terms denote the intra-orbital, inter-orbital Hubbard repulsion and the Hund's rule coupling as well as the pair hopping. The spacial rotation symmetry requires $U=V+2J_H$. Here we let $U$ and $J_H/U$ to be tuning parameters and study the parameter dependence of the results.

\begin{figure}[htbp]
\centering
\includegraphics[width=0.4\textwidth]{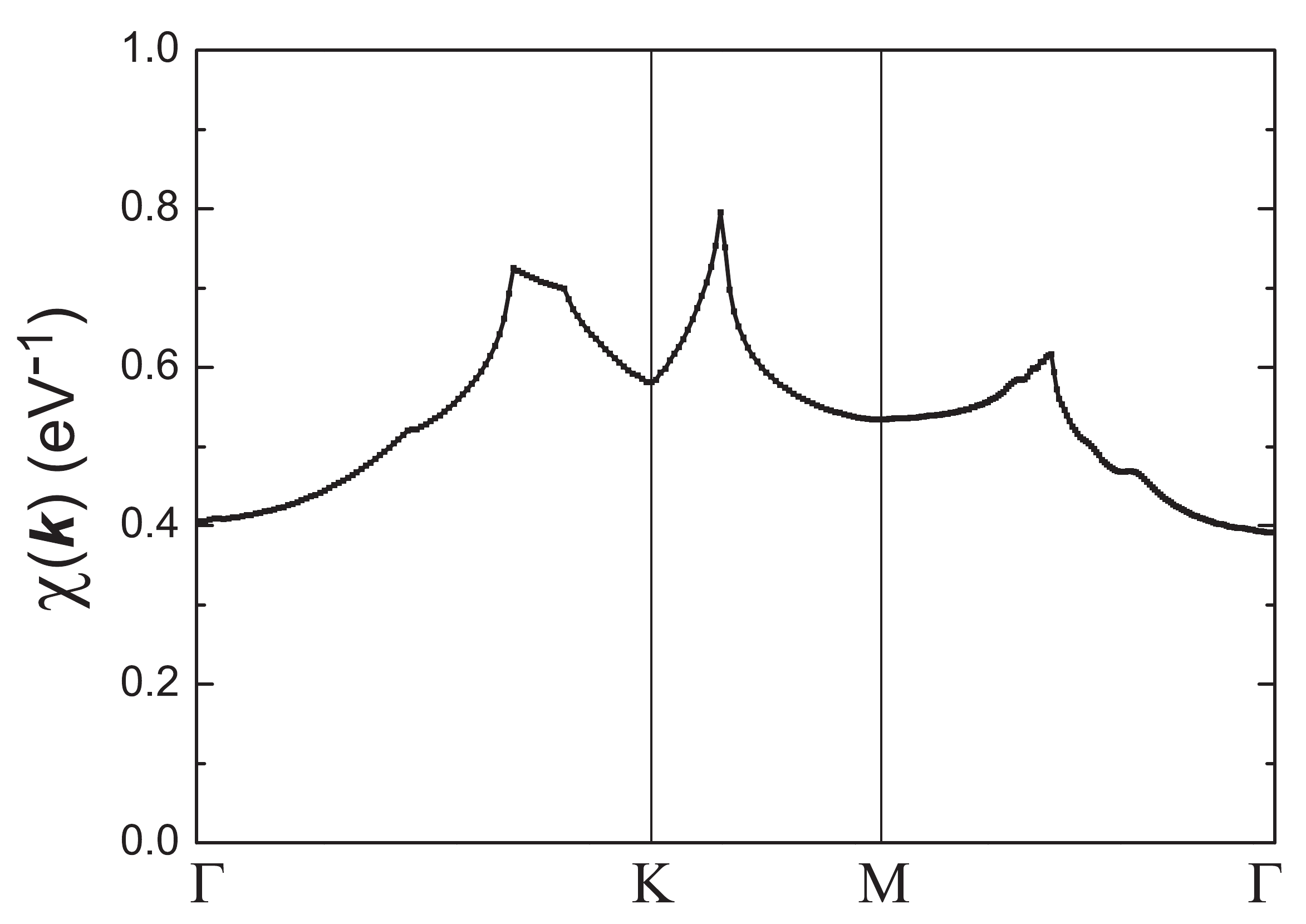}
\caption{The $\bm{k}$-space distribution along the high-symmetric lines of the largest eigenvalue of zero temperature susceptibility matrix $\chi^{(0)}_{pq}(\bm{k})\equiv\chi^{(0)pp}_{qq}(\bm{k},i\omega_n=0)$ for the undoped system.}\label{chi}
\end{figure}

According to the standard multi-orbital RPA approach \cite{RPA1,RPA2,RPA3,Kuroki,Scalapino1,Scalapino2,Liu2013,Wu2014,Ma2014,Zhang2015}, we first define the following bare susceptibility for the non-interacting case ($U=V=J_H=0$):
\begin{align}\label{chi0}
\chi^{(0)pq}_{st}(\bm{k},\tau)\equiv
&\frac{1}{N}\sum_{\bm{k}_1\bm{k}_2}\left\langle
T_{\tau}c_{p}^{\dagger}(\bm{k}_1,\tau)
c_{q}(\bm{k}_1+\bm{k},\tau)\right.                      \nonumber\\
&\left.\times c_{s}^{\dagger}(\bm{k}_2+\bm{k},0)
c_{t}(\bm{k}_2,0)\right\rangle_0,
\end{align}
Here $\langle\cdots\rangle_0$ denotes the thermal average for the noninteracting system, $T_{\tau}$ denotes the time-ordered product, and $p,q,s,t=1,2,3$ are the orbital indices. Fourier transformed to the imaginary frequency space, the bare susceptibility can be expressed by the following explicit formulism:
\begin{align}\label{chi0e}
\chi^{(0)pq}_{st}&(\bm{k},i\omega_n)
=\frac{1}{N}\sum_{\bm{k}'\alpha\beta}
\xi^{\alpha}_{t}(\bm{k}')
\xi^{\alpha*}_{p}(\bm{k}')
\xi^{\beta}_{q}(\bm{k}'+\bm{k})                         \nonumber\\
&\times\xi^{\beta*}_{s}(\bm{k}'+\bm{k})
\frac{f(\varepsilon^{\beta}_{\bm{k}'+\bm{k}}-\mu_c)
-f(\varepsilon^{\alpha}_{\bm{k}'}-\mu_c)}
{i\omega_n+\varepsilon^{\alpha}_{\bm{k}'}
-\varepsilon^{\beta}_{\bm{k}'+\bm{k}}},
\end{align}
where $\alpha,\beta=1,2,3$ are band indices, $\varepsilon^{\alpha}_{\bm{k}}$ and $\xi^{\alpha}_{l}\left(\bm{k}\right)$ are the $\alpha$-th eigenvalue and eigenvector
of the $h(\bm{k})$ matrix, respectively, and $f$ is the Fermi-Dirac distribution
function. In Fig. \ref{chi}, we show the $\bm{k}$-dependence of the largest eigenvalue of zero temperature susceptibility matrix $\chi^{(0)pp}_{ss}(\bm{k},i\omega_n=0)$ along the high symmetry lines in the Brillouin Zone. Clearly, the largest eigenvalue peaks around the K-points and far away from the $\Gamma$-point, which implies that the dominating spin correlation of the system is AFM and will mediate singlet pairing. This susceptibility peak originates from the good FS nesting between opposite edges of the nearly hexagonal FS, as show in Fig.\ref{bandfsdos}(b).

When the interactions in Eq. (\ref{model}) are turned on, we can further define the spin and charge susceptibilities. In the RPA level, repulsive Hubbard interactions suppress the charge susceptibility and enhance the spin susceptibility \cite{RPA1,RPA2,RPA3,Kuroki,Scalapino1,Scalapino2,Liu2013,Wu2014,Ma2014,Zhang2015}. Note that there is a critical interaction strength $U_c$ which depends on the ratio $J_H/U$. For $U$=$U_c$, the spin susceptibility diverges there, which invalidates the RPA treatment and implies the formation of long-range magnetic order. When the interaction strength $U<U_c$, there can be short-ranged spin and charge fluctuations in the system. Through exchanging these fluctuations between a Cooper pair, exotic superconducting states will emerge in the system. If we consider the scattering of a Cooper pair from the state $(\bm{k}',-\bm{k}')$ in the $\beta$-th band to the state $(\bm{k},-\bm{k})$ in the $\alpha$-th band via exchanging spin or charge fluctuations, we can obtain the effective interaction vertex $V^{\alpha\beta}(\bm{k},\bm{k}')$, and the corresponding linearized gap equation near the superconducting critical temperature $T_c$ \cite{Scalapino1,Scalapino2}:
\begin{align}\label{gapeq}
-\frac{1}{(2\pi)^2}\sum_{\beta}\oint_{FS}
d^{2}\bm{k}'_{\Vert}\frac{V^{\alpha\beta}(\bm{k},\bm{k}')}
{v^{\beta}_{F}(\bm{k}')}\Delta_{\beta}(\bm{k}')=\lambda
\Delta_{\alpha}(\bm{k}).
\end{align}
Here the integration is along various FS patches labelled by $\alpha$ or $\beta$, $v^{\beta}_F(\bm{k}')$ is the Fermi velocity, and $\bm{k}'_\parallel$ is the component of $\bm{k}'$ along the FS. Solving this gap equation as an eigenvalue problem, one obtains each pairing eigenvalue $\lambda$ and the corresponding normalized eigenvector $\Delta_\alpha(\bm{k})$ as the relative pairing gap function. The leading pairing symmetry is determined by the $\Delta_\alpha(\bm{k})$ corresponding to the largest $\lambda$. The critical temperature $T_c$ is determined by $\lambda$ through $T_c=$ cutoff energy $\cdot~ e^{-1/\lambda}$, where the cutoff energy scales with the low energy bandwidth.

\begin{figure}[htbp]
\centering
\includegraphics[width=0.48\textwidth]{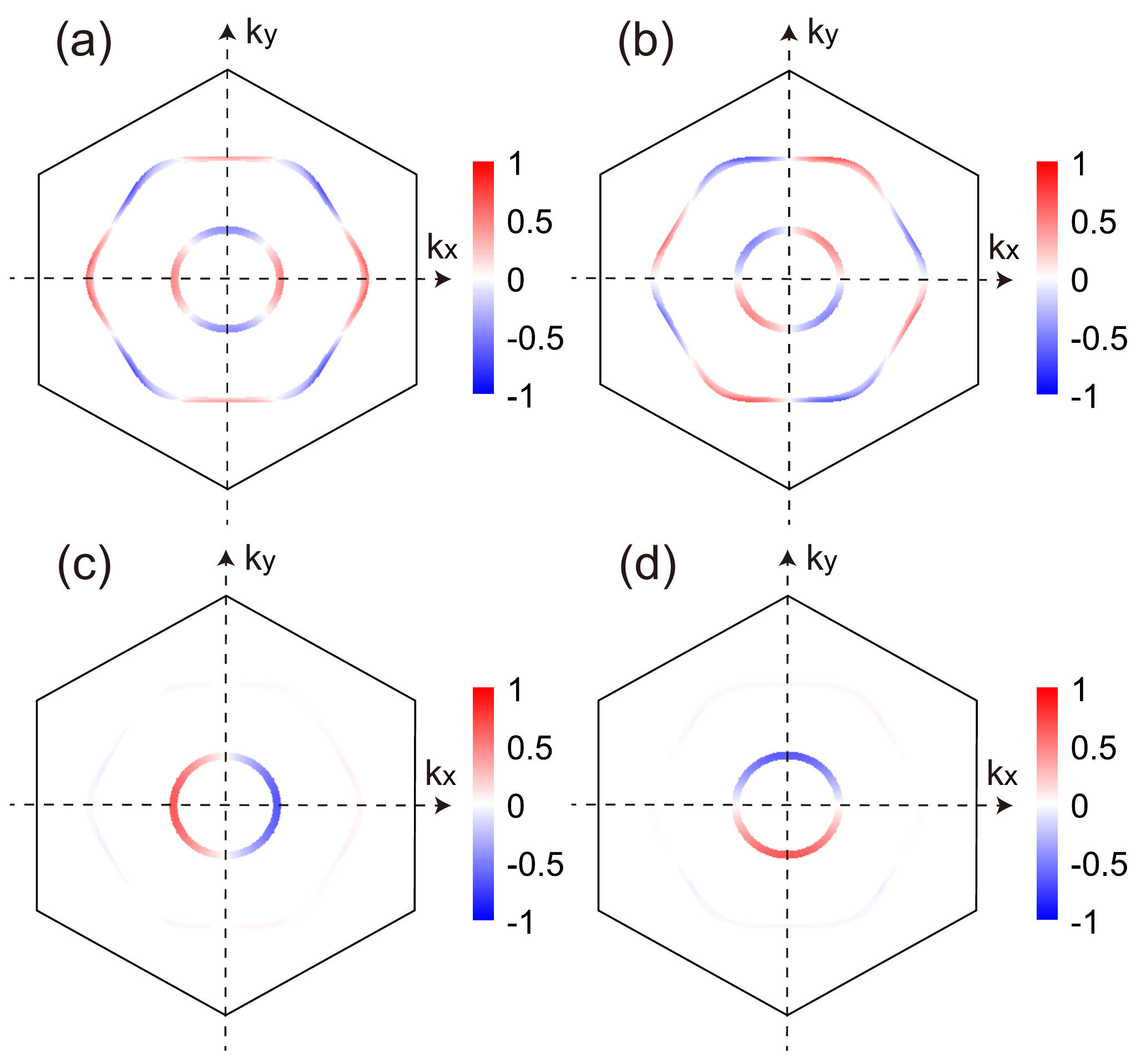}
\caption{The distribution of the relative gap function of (a) $d_{x^2-y^2}$ and (b) $d_{xy}$ symmetries at $U=1$eV and $J_H=0$, as well as those of (c) $p_x$ and (d) $p_y$ symmetries at $U=0.5$eV and $J_H=0.5U$ for the undoped system.}\label{gap}
\end{figure}

Corresponding to the $D_{3h}$ point group of YNiO$_3$, the possible superconducting pairing symmetries include $s$-, $p$-, $d$-, and $f$-wave ones. The eigenvector(s) $\Delta_{\alpha}(\bm{k})$ for each eigenvalue $\lambda$ obtained from gap equation (\ref{gapeq}) as the basis function(s) forms an irreducible representation of the $D_{3h}$ point group. While the $s$ and $f$ symmetries each forms a 1D representation of the point group with nondegenerate pairing eigenvalues, the $(d_{x^2-y^2},d_{xy})$ and $(p_x,p_y)$ symmetries each forms a 2D representation with degenerate pairing eigenvalues. The gap function of the $d_{x^2-y^2}$ and $d_{xy}$ symmetries are symmetric and antisymmetric about both the $x$ and $y$ axes respectively as shown in Figs. \ref{gap}(a) and \ref{gap}(b). The gap function of the $p_x$($p_y$) symmetry is symmetric about the $x(y)$ axis and antisymmetric about the $y(x)$ axis as shown in Fig. \ref{gap}(c)(\ref{gap}(d)).

\begin{figure}[htbp]
\centering
\includegraphics[width=0.48\textwidth]{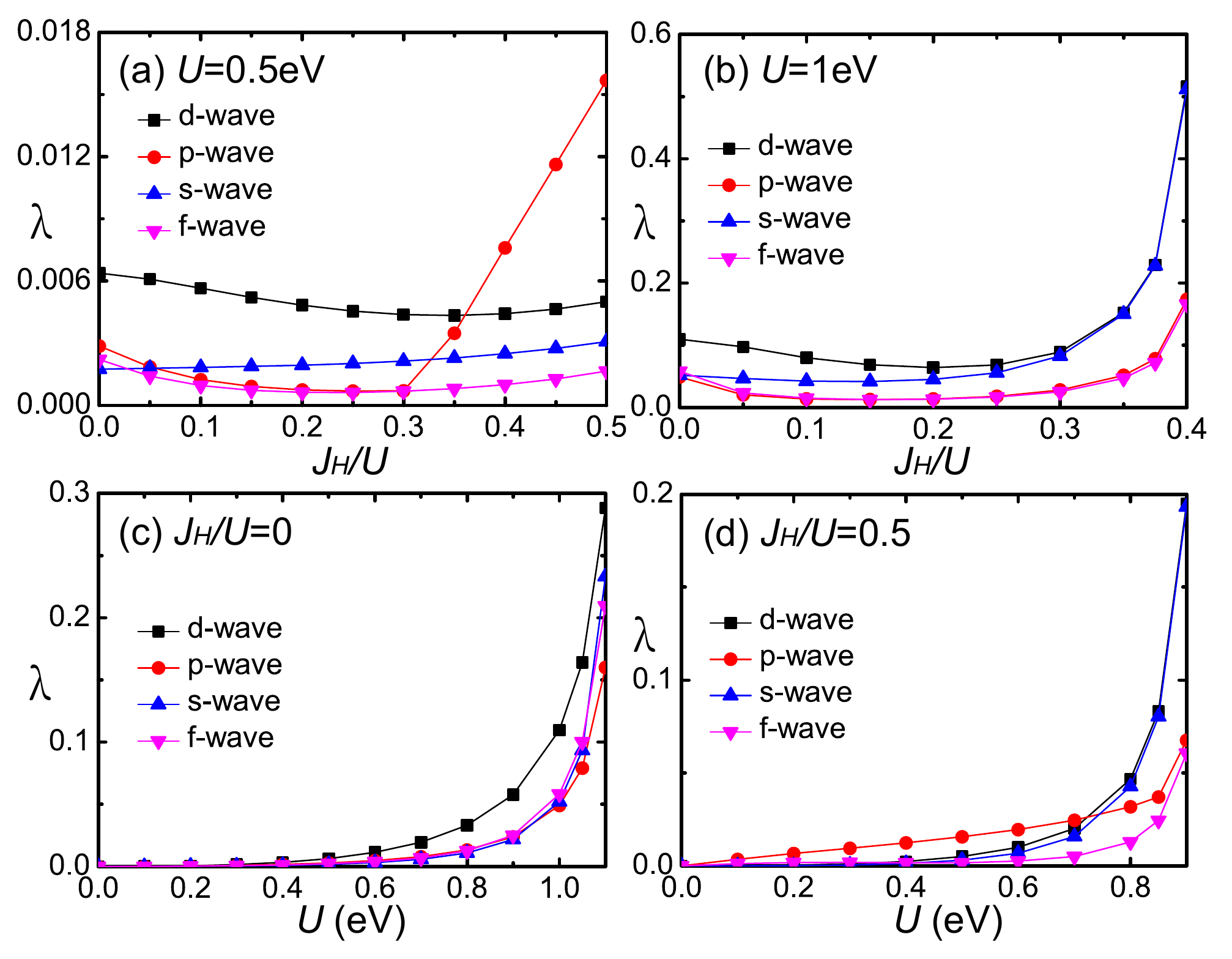}
\caption{The $J_H/U$- and $U$-dependence of the largest eigenvalues $\lambda$ for all pairing symmetries of the undoped system. (a) $U=0.5$eV, (b) $U=1$eV, (c) $J_H/U=0$, (d) $J_H/U=0.5$.}\label{pair}
\end{figure}

The $J_H/U$-dependence of the largest eigenvalues $\lambda$ for all pairing symmetries is shown in Figs. \ref{pair}(a) and \ref{pair}(b). Clearly, when $U=0.5$eV, $d$- and $p$-wave symmetries dominate for small and large values of $J_H/U$, respectively. When $U=1$eV, $d$-wave symmetry dominates for all values of $J_H/U$ with $s$-wave one as close candidate for large values of $J_H/U$. The $U$-dependence of the largest eigenvalues $\lambda$ for all pairing symmetries is shown in Figs. \ref{pair}(c) and \ref{pair}(d). Clearly, when $J_H/U=0$, $d$-wave symmetry dominates for all values of $U$. When $J_H/U=0.5$, $p$- and $d$-wave symmetries dominate for small and large values of $U$, respectively, with $s$-wave one as close candidate for large values of $U$. Physically, the $d$-wave SC at weak $J_H/U$ ($<0.35$) is driven by the AFM spin fluctuations revealed in Fig. \ref{chi}, and the $p$-wave SC at strong $J_H/U$ ($>0.35$) is driven by the Hund¡¯s rule coupling (similar Hund's rule coupling driven on-site inter-orbital odd-parity pairing was also proposed in other systems \cite{KCrAs,BiH}). For realistic parameters for Fe, the d-wave is the leading pairing symmetry. Among other pairing channels, the $d_{x^2-y^2}$ and $d_{xy}$ pairing gap functions shown in Figs. \ref{gap}(a) and \ref{gap}(b) mostly satisfy the condition that the gap functions connected by the nesting vector $\bm{Q}$ must have a sign change.

Since the $d_{x^2-y^2}$ and $d_{xy}$ pairing states are degenerate, they will mix to lower the energy below the critical temperature $T_c$. To determine this mixture, we set $\Delta^{\alpha}_{\bm{k}}=K_1d^{\alpha}_{x^2-y^2}(\bm{k})+(K_2+iK_3)d^{\alpha}_{xy}(\bm{k})$, where $d^{\alpha}_{x^2-y^2}(\bm{k})$ and $d^{\alpha}_{xy}(\bm{k})$ denote the normalized gap functions of corresponding symmetries, and $K_1$, $K_2$, and $K_3$ are the mixing coefficients. Our energy minimization gives $K_1=\pm K_3$ and $K_2=0$, which leads to the fully-gapped $d_{x^2-y^2}\pm i d_{xy}$ (abbreviated as $d+id$) SC. This mixture of the two $d$-wave pairings satisfies the requirement that the gap nodes should avoid the FS to lower the energy. Physically, the singlet $d+id$ pairing is mediated by the strong AFM spin fluctuations revealed by Fig. \ref{chi}. Similarly, one can verify that the degenerate $p_x$ and $p_y$ pairing states will also mix into the fully-gapped $p_x\pm ip_y$ (abbreviated as $p+ip$) pairing state to lower the energy below $T_c$.

\begin{figure}[htbp]
\centering
\includegraphics[width=0.4\textwidth]{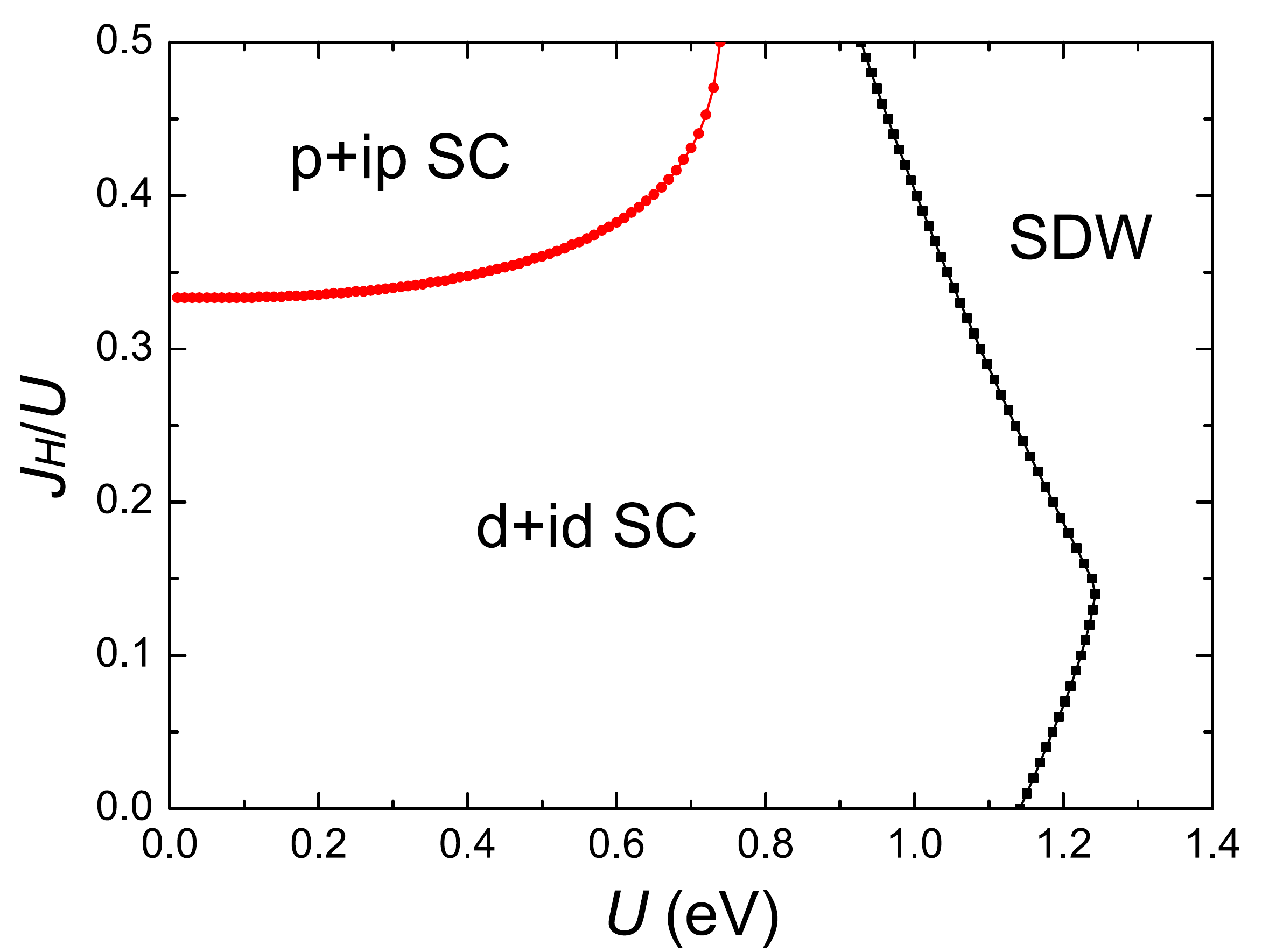}
\caption{The ground state phase diagram of the undoped system in $U-J_H/U$ plane.}\label{phase}
\end{figure}

The ground state phase diagram in $U-J_H/U$ plane is shown in Fig. \ref{phase}, where three possible phases are present. For $U>U_c$, which is around $1\sim1.2$eV and is $J_H/U$-dependent, the SDW order emerges. For $U<U_c$, the $d+id$ and $p+ip$ pairings states emerge, which are separated by a $U$-dependent critical value of $J_H/U$. For $J_H/U$ below or above this critical value, the $d+id$ or $p+ip$ wave pairing is the leading pairing symmetry, respectively. In the limit of $U\to 0$, this critical value of $J_H/U$ tends to $1/3$. with the enhancement of $U$, this critical value tends to $0.5$ around $U=0.75$eV.

\begin{figure}[htbp]
\centering
\includegraphics[width=0.48\textwidth]{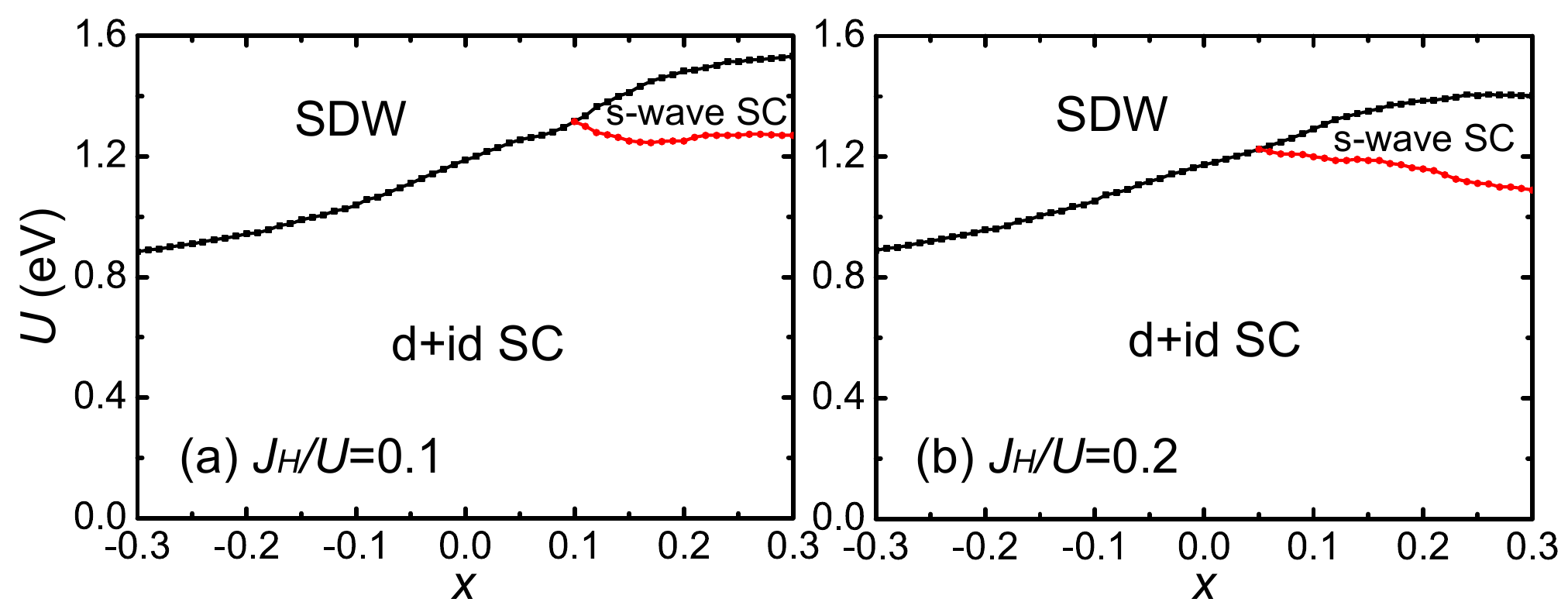}
\caption{The ground state phase diagrams in $x-U$ plane for realistic values of $J_H/U$, where $x$ denotes the doping level. (a) $J_H/U=0.1$ and (b) $J_H/U=0.2$.}\label{doping}
\end{figure}

To investigate the robustness of the $d+id$ pairing state against the doping, we further study the pairing symmetry of the system at different doping levels. Defining the doping level $x=n_e-3$, where $n_e$ is the electron number per site in our model, we show the ground state phase diagram of the system in $x-U$ plane for realistic $J_H/U$ in Fig. \ref{doping}. Clearly, the $d+id$ pairing state serves as the ground state of the system across the doping range $x=-0.3\sim0.3$ when $U$ is smaller than $U_c$ or a critical $U$ separating the $d+id$ pairing state and the $s$-wave pairing state which emerges for large $x$ and $U$. Noting that such a doping range is almost wide enough to cover the experimentally accessible doping levels, we can conclude that the $d+id$ SC is robust against doping.

\section{Strong coupling limit: Mean field approach}
In the strong-coupling limit, we consider superexchange type of interactions. From DFT calculations on YNiO$_3$ \cite{YNO}, in our three-orbital model, the dominant hopping parameter is the nearest-neighbor (NN) hopping, which provides us the following exchange interaction term:
\begin{align}\label{modelFM}
H_{J}=&J\sum_{\langle ij\rangle \mu\nu}\bm{S}_{i\mu}\cdot\bm{S}_{j\nu},
\end{align}
with $J>0$, representing AFM superexchange interaction suggested by DFT calculations. Here $\mu,~\nu\in\{d_{z^2}, d_{xy}, d_{x^2-y^2}\}$, $\langle ij\rangle$ denotes NN bond along the three directions $(\bm{e}_1,\bm{e}_2,\bm{e}_3)$ in our two-dimension model, and $\bm{S}_{\mu i}=\frac{1}{2}\sum_{\sigma\sigma'}c^{\dagger}_{ i\mu \sigma}\bm{\sigma}_{\sigma\sigma'}c_{ i\mu \sigma'}$ is the local spin operator for the $\mu$-th orbital. Adding this interaction term into our TB model, we obtain an effective $t-J$ model:
\begin{align}\label{modelTJ}
H=H_{tb}+H_{J},
\end{align}

For the purpose of determining the pairing symmetry, we shall omit the no-double-occupance constraint for simplicity and perform a mean-field analysis on the model. Due to the AFM superexchange interactions, the interaction Hamiltonian can be mean-field decoupled in the singlet channel as,
\begin{align}\label{HDC}
H_{J}=&\sqrt{2}\left(\sum_{\langle ij\rangle\mu\nu}\Delta_{i\mu j\nu}\bm{P}^{\dagger}_{i\mu j\nu}+H.c.\right)+\frac{8J}{3}\sum_{\langle ij\rangle\mu\nu}\left| \Delta_{i\mu j\nu}\right|^2.
\end{align}
Here $\bm{P}^{\dagger}_{i\mu j\nu}=\frac{1}{\sqrt{2}}(c^{\dagger}_{i\mu\uparrow}c^{\dagger}_{j\nu\downarrow}-
c^{\dagger}_{i\mu\downarrow}c^{\dagger}_{j\nu\uparrow})$ is the spin-singlet pairing operator and  $\Delta_{i\mu j\nu}=\frac{-3J}{4\sqrt{2}}\langle\bm{P}_{i\mu j\nu}\rangle$ is the pairing order parameters.

Together with the tight-binding part, the total mean-field Hamiltonian in the momentum space is
\begin{align}\label{HDC}
H_{mf}=&\sum_{\bm{k}}\mathbf{\Psi}^{\dagger}(\bm{k})H(\bm{k})\mathbf{\Psi}(\bm{k})+\frac{8J}{3}\sum_{\langle ij\rangle \mu\nu}\left | \Delta_{ij\mu\nu} \right |^2,
\end{align}
with
\begin{align}
H(\bm{k})=&\begin{pmatrix}
h(\bm{k})-\mu_cI & \Delta(\bm{k})     \\
\Delta^{\dagger}(\bm{k}) & -h^{*}(\bm{-k})+\mu_cI
\end{pmatrix},              \nonumber\\
\Delta(\bm{k})=&
\begin{pmatrix}
\Delta_{11}(\bm{k})&\Delta_{12}(\bm{k})& \Delta_{13}(\bm{k})\\
\Delta_{21}(\bm{k})&\Delta_{22}(\bm{k})& \Delta_{23}(\bm{k})\\
\Delta_{31}(\bm{k})&\Delta_{32}(\bm{k})& \Delta_{33}(\bm{k})\\
\end{pmatrix},
\end{align}
where $\mathbf{\Psi}^{\dagger}(\bm{k})=[c^{\dagger}_{\bm{k}1\uparrow},c^{\dagger}_{\bm{k}2\uparrow},c^{\dagger}_{\bm{k}3\uparrow},
c_{\bm{-k}1\downarrow},c_{\bm{-k}2\downarrow},c_{\bm{-k}3\downarrow}]$ and $\Delta_{\mu\nu}(\bm{k})=\sum_{l=1,3}[\Delta_{i\mu ,i+\bm{e}_{l}\nu}e^{i\bm{k}\cdot \bm{e}_{l}}+\Delta_{i\nu, i+\bm{e}_{l}\mu}e^{-i\bm{k}\cdot \bm{e}_{l}}]$. Note that a constant $\sum_{\bm{k}\mu}[h_{\mu\mu}(\bm{k})-\mu_c]$ has been neglected here since it has no contribution to the dynamics.
The above mean-field Hamiltonian can be solved via diagonalizing $H(\bm{k})$ by a unitary transformation: $U^{\dagger}(\bm{k})H(\bm{k})U(\bm{k})=Diag(E_n(\bm{k}))$, with the eigenvalues $E_n(\bm{k})=-E_{n+3}(\bm{k})>0$ $(n=1,2,3)$. The self-consistent gap equations are
\begin{align}\label{SCE}
\Delta_{i\mu,i+\bm{e}_{l}\nu}=&\frac{-3J}{8N}
\sum_{\bm{k}n}f(E_n(\bm{k}))
\left[e^{-i\bm{k}\cdot\bm{e}_{l}}
U^{*}_{\nu+3,n}(\bm{k})U_{\mu,n}(\bm{k})\right.    \nonumber\\
&\left.~~~~~~~~+e^{i\bm{k}\cdot \bm{e}_{l}} U^{*}_{\mu+3,n}(\bm{k})U_{\nu,n}(\bm{k})\right],
\end{align}
where $f(E_n(\bm{k}))$ is the Fermi-Dirac distribution function, and here we only consider the case of zero temperature. Solving these gap equations, we get the 27 complex gap amplitudes $\Delta_{i\mu,i+\bm{e}_{l}\nu}~(\mu/\nu/l=1,2,3)$.
\begin{figure}[htbp]
\centering
\includegraphics[width=0.48\textwidth]{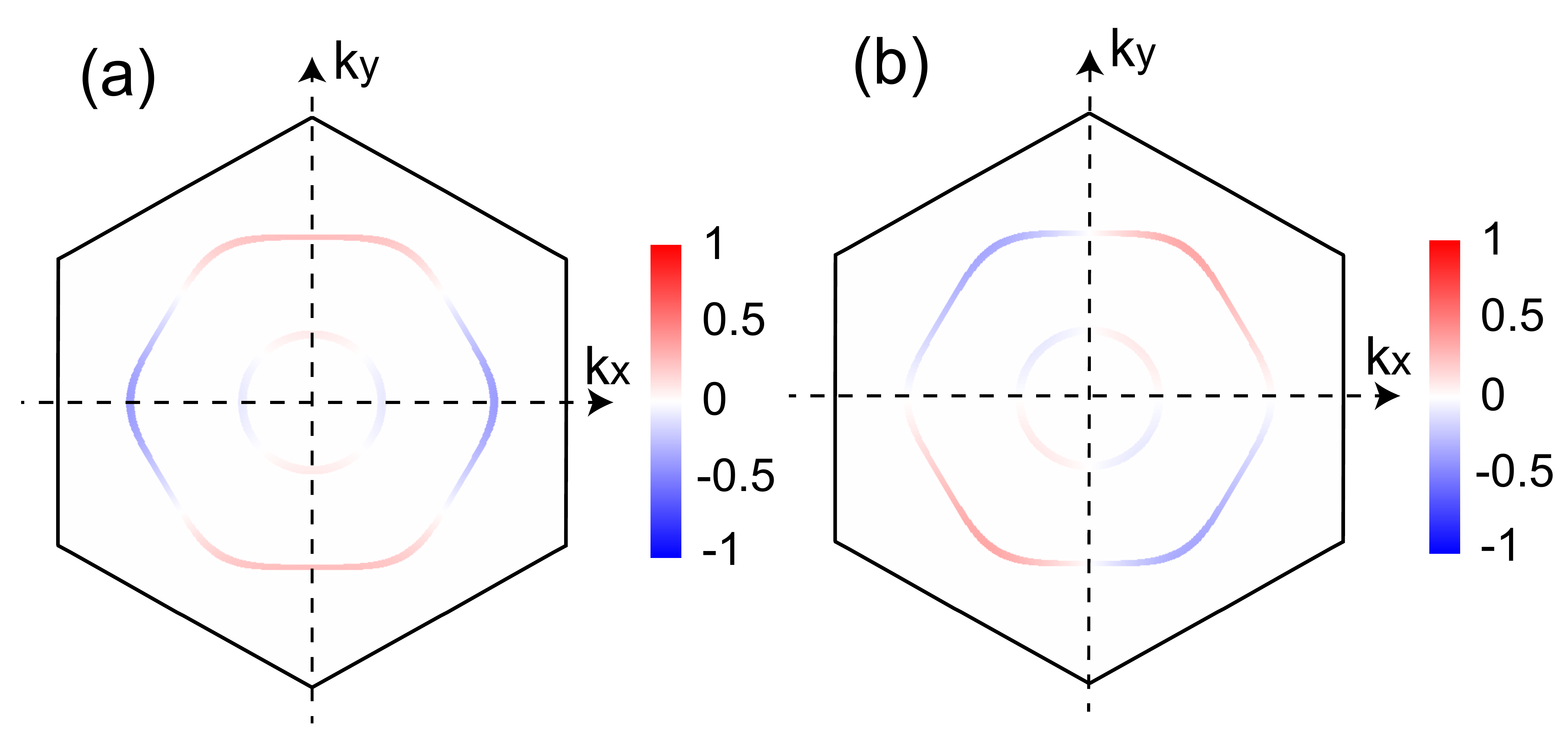}
\caption{The distribution of the (a) real and (b) imaginary parts of the gap function over the FSs at $J=0.2$eV for the undoped system.}\label{d+id}
\end{figure}
Our mean-field calculations for the above self-consistent gap equations always yield doubly-degenerate solutions within the wide doping range $x=-0.3\sim 0.3$ for realistic superexchange parameter $J$. Further investigation into the two doubly-degenerate solutions suggests that time-reversal symmetry is broken in both states, and the two degenerate states are time-reversal related. In order not to be disturbed by the too many (totally 27) gap amplitudes and enhance the visibility about the pairing gap symmetry, we project the gap function onto the FS within the intra-band pairing approximation via
\begin{align}\label{projection}
&\sum_{\mu\nu}c^{\dagger}_{\bm{k}\mu\uparrow} c^{\dagger}_{-\bm{k}\nu\downarrow}\Delta_{\mu\nu}(\bm{k})\nonumber\\\to&
\sum_{\mu\nu\alpha}c^{\dagger}_{\bm{k}\alpha\uparrow} c^{\dagger}_{-\bm{k}\alpha\downarrow}\xi^{\alpha*}_{\mu}\bm(k)
\xi^{\alpha*}_{\nu}\bm(-k)\Delta_{\mu\nu}(\bm{k})\nonumber\\\equiv& \sum_{\alpha}c^{\dagger}_{\bm{k}\alpha\uparrow} c^{\dagger}_{-\bm{k}\alpha\downarrow}\tilde{\Delta}^{\alpha}(\bm{k}).
\end{align}

Figure \ref{d+id} shows the distribution of the relative (normalized) gap function on the FSs for one of the doubly-degenerate solutions of the self-consistent gap equation (\ref{SCE}) for a typical superexchange interaction parameter $J=0.2$eV in the undoped system. The real and imaginary part of the complex gap function are shown in Fig. \ref{d+id}(a) and \ref{d+id}(b), respectively. Clearly, the pairing symmetry shown here in Fig. \ref{d+id} is consistent with the pairing symmetry shown in Fig. \ref{gap}(a) and \ref{gap}(b), which is $d_{x^2-y^2}+id_{xy}$.  The other solution has the same real part and the same but one minus sign different imaginary part as that shown in Fig. \ref{gap}(a) and \ref{gap}(b), which is $d_{x^2-y^2}-id_{xy}$. Therefore, both weak-coupling and strong-coupling calculations consistently yield the $d+id$ as the leading pairing symmetry for a wide doping range and realistic interaction parameters.

\section{Topological SC}
It's known that the $d+id$ chiral SC can be topologically nontrivial \cite{Read_and_Green}. To investigate the topological properties of the $d+id$ SC proposed here, we calculate the following topological invariant Chern number \cite{ChernN1,ChernN2}:
\begin{align}\label{chern}
C_1=&\int_{BZ}d^2\bm{k}\sum_{mn}[f(E_m(\bm{k}))-f(E_n(\bm{k}))]           \nonumber\\
&\times\frac{u^{\dag}_m(\bm{k})\partial_{k_x}H(\bm{k})u_n(\bm{k})
u^{\dag}_n(\bm{k})\partial_{k_y}H(\bm{k})u^{\dag}_m(\bm{k})}
{\left[E_m(\bm{k})-E_n(\bm{k})\right]^2},
\end{align}
where $E_n(\bm{k})$ and $u_n(\bm{k})$ are the eigenvalues and eigenstates of $H(\bm{k})$, respectively. We calculated the Chern number of the system within the doping range $x=-0.3\sim 0.3$, and obtained a constant value $8$. The Chern number $8$ obtained here is the consequence of the three combined factors, i.e., two electron pockets, two spin species, and angular momentum $2$ for the pairing, with $2\times 2\times 2=8$.

\begin{figure}[htbp]
\centering
\includegraphics[width=0.40\textwidth]{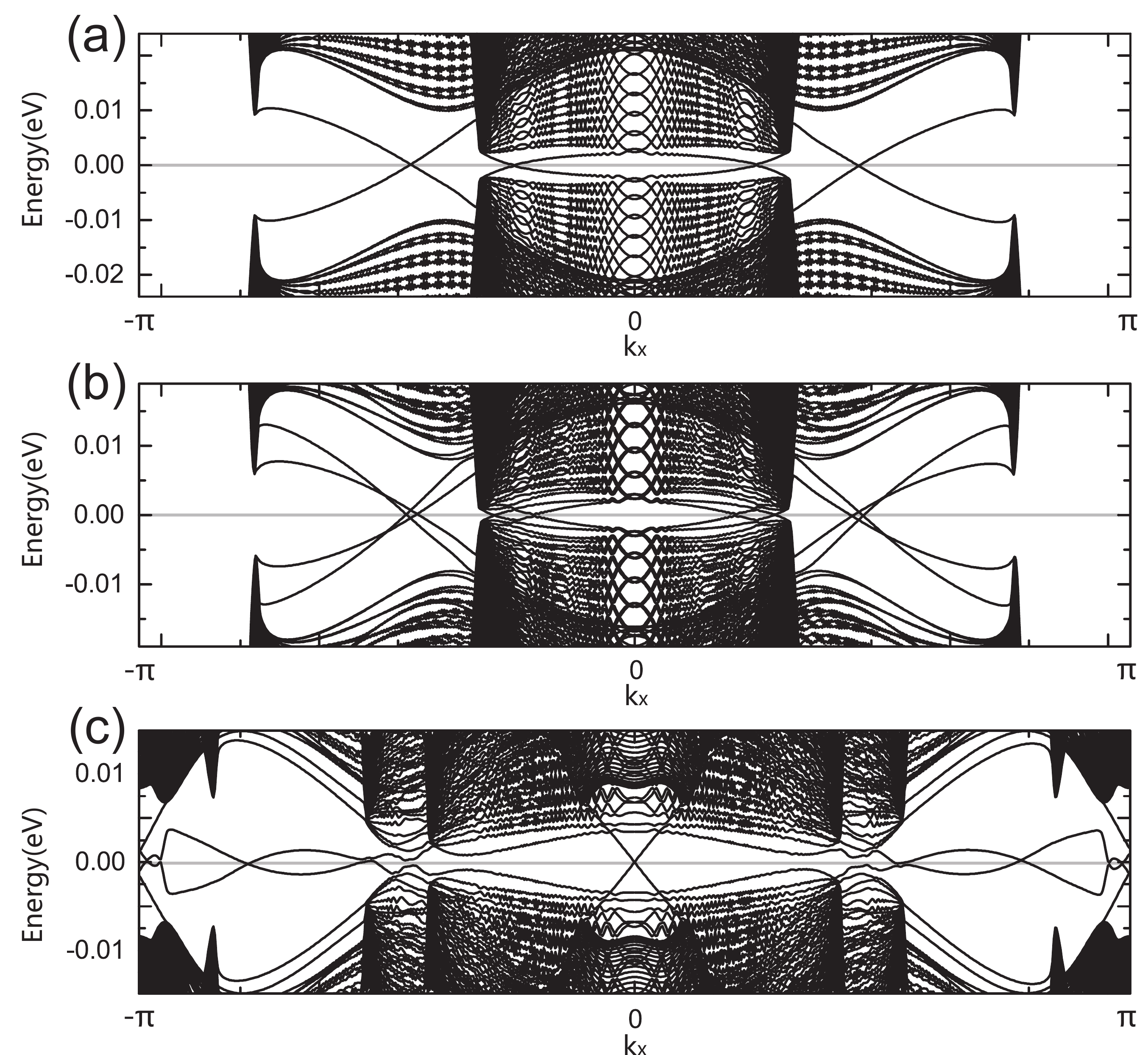}
\caption{The edge spectra of the $d+id$ chiral superconducting states for (a) $\lambda_{R}=0, V_{z}=0,\mu=3.0$eV, (b) $\lambda_{R}=5$meV, $V_{z}=0,\mu=3.0$eV, (c) $\lambda_{R}=30$meV, $V_{z}=35$meV, $\mu=2.755$eV.}\label{ep}
\end{figure}

The topological property of the system can also be manifested by its edge spectrum. Here we adopt periodic boundary condition along one direction (defined as $\bm{x}$), and open boundary condition along the other direction, which is $120^{o}$ rotated from $\bm{x}$.  Using the 27 complex gap parameters $\Delta_{i\mu,i+\bm{e}_{l}\nu}~(\mu/\nu/l=1,2,3)$ obtained from the mean-field solution, we write the mean-field BdG Hamiltonian in the real space with such an open-periodic boundary condition, and show its energy spectrum in Fig. \ref{ep}(a). Figure. \ref{ep}(a) seems to illustrate four branches of chiral gapless Dirac modes on each edge. However, each branch is doubly degenerate due to the spin degeneracy.

Such spin degeneracy can be lifted up by a small Rashba SOC added in the formulism of
\begin{align}\label{Rashba}
H_{R}=i\lambda_R\sum_{<ij>}c^{\dagger}_i(\bm{\sigma}\times\bm{d}_{ij})_zc_j,
\end{align}
where the coupling constant $\lambda_R=5$meV, the vector $\bm{d}_{ij}$ points from $j$ to $i$, and $c^{\dagger}_i=(c^{\dagger}_{i\uparrow},c^{\dagger}_{i\downarrow})$.
Such an extra SOC term does not change the Chern-number, but it induces spin splitting in the edge spectrum, as shown in Fig. \ref{ep}(b). Consequently, there are eight branches of chiral gapless modes on each edge.

Therefore, the $d+id$ SC obtained here in YNiO$_3$ is chiral topological SC with a high Chern-number $8$, which hosts $8$ gapless chiral modes on each edge. These chiral modes are topologically protected and are robust against perturbations. If we place the superconducting system in an external magnetic field between its lower and upper critical field, the Zeeman coupling with the magnetic field could change the Chern-number to odd integers and enable the vortex core to host a single Majorana zero mode at $k_x=0$, when the field strength is properly tuned. Such vortex core can be used to equip topological quantum computation. To verify this point, we add the following Zeeman term into the Hamiltonian,
\begin{eqnarray}
H_{z}=V_z\sum_{i,\mu}\left(c^{\dagger }_{i\mu\uparrow}c_{i\mu \uparrow }-c^{\dagger }_{i\mu\downarrow }c_{i\mu\downarrow }\right).
\end{eqnarray}
For our parameters chosen as $\lambda_R=30$meV, $\mu_c=2.755$eV and $V_z=35$meV, the Chern-number calculated by Eq.(\ref{chern}) yield 1. As a result, in the edge spectrum shown in Fig.\ref{ep}(c), one verifies a zero energy cross at $k_x=0$, which is identified as the Majorana zero mode. Note that there also exist several occasional crosses in the spectrum, which however is not topologically protected.

\section{Conclusion and discussion}
In conclusion, we have studied the pairing symmetry of the newly predicted system that can host high-$T_c$ superconductivity. Starting from its effective three-orbital TB model, the system consistently yields $d+id$ chiral SC as the leading pairing symmetry within wide doping range for realistic interaction parameters in our combined weak and strong coupling methods.   This superconducting state breaks the time-reversal symmetry and is topologically nontrivial with a high Chern number $8$ .  With a weak Rashba SOC in the system, the vortex cores under magnetic field can carry Majorana zero mode, which can be used to equip quantum computation.

For realistic material, the undoped system might be AFM Mott-insulator, instead of superconductor. The reason is that for the three active orbitals near the Fermi level, the band filling of three electrons per unit cell just makes the half-filling case. In the strong coupling limit at half-filling, the no-double-occupance constraint omitted here would drive the system into AFM Mott-insulator. However, when extra electrons or holes are doped into the system via atomic substitution, the $d+id$ chiral SC would be realized in real materials with high superconducting critical temperature. Such high-$T_c$ topological SC would be intriguing and of fundamental importance.

\section*{Acknowledgements}
 This work is supported by the NSFC (Grant Nos. 11674025, 11604013, 11334012, 11274041, 1190020, 11534014), the Ministry of Science and Technology of China 973 program (No. 2015CB921300, 2017YFA0303100), and Beijing Natural Science Foundation (Grant No. 1174019).

\end{document}